\documentclass[fleqn,usenatbib]{mnras}

\usepackage{newtxtext,newtxmath}

\usepackage[T1]{fontenc}

\DeclareRobustCommand{\VAN}[3]{#2}
\let\VANthebibliography\thebibliography
\def\thebibliography{\DeclareRobustCommand{\VAN}[3]{##3}\VANthebibliography}

\usepackage{graphicx}	
\usepackage{amsmath}	
\usepackage{amssymb}	
\usepackage{subfigure}

\title[SFMS1.6v8]{The H\boldmath$\alpha$ star formation main sequence in cluster and field galaxies at $z\sim1.6$}
\author[Nantais et al.]{Julie Nantais$^{1}$,\thanks{E-mail: julie.nantais@unab.cl}
	Gillian Wilson$^{2}$,
	Adam Muzzin$^{3}$,
    Lyndsay J. Old$^{4}$,
    \newauthor
	Ricardo Demarco$^{5}$,
    Pierluigi Cerulo$^{5}$,
    Michael Balogh$^{6,7}$,
    Gregory Rudnick$^{8}$,
    \newauthor
    Jeffrey Chan$^{2}$,
    M. C. Cooper$^{9}$,  
    Ben Forrest$^{2}$,
    Brian Hayden$^{10,11}$,
    \newauthor
    Chris Lidman$^{12,13}$,
    Allison Noble$^{14}$,
    Saul Perlmutter$^{10,11}$,
    Carter Rhea$^{15}$,
    \newauthor
    Jason Surace$^{16}$,
    Remco van der Burg$^{17}$,
    and Eelco van Kampen$^{17}$\\
	$^{1}$Departamento de Ciencias F\'isicas, Universidad Andr\'es Bello, Fernandez Concha 700, Las Condes 7591538, RM, Chile\\
    $^{2}$Department of Physics and Astronomy, University of California at Riverside, 900 University Avenue, Riverside, CA 92521, USA \\
    $^{3}$Department of Physics and Astronomy, York University, 7400 Keele St., Toronto, Ontario MJ3 1P3, Canada \\
    $^{4}$European Space Agency, European Space Astronomy Center, Villanueva de la Ca\~nada, E-2691, Madrid, Spain\\
    $^{5}$Departamento de Astronomía, Universidad de Concepción, Concepci\'on, Chile \\
    $^{6}$Department of Physics and Astronomy, University of Waterloo, Waterloo, Ontario N2L 3G1, Canada\\
    $^{7}$Waterloo Centre for Astrophysics, University of Waterloo, Waterloo, Ontario, N2L 3G1, Canada\\
	$^{8}$Department of Physics and Astronomy, University of Kansas, 1251 Wescoe Hall Drive, Lawrence, KS 66045, USA \\
	$^{9}$Department of Physics and Astronomy, University of California at Irvine, 4129 Frederick Reines Hall, Irvine, CA 92697, USA\\
	$^{10}$Physics Division, Lawrence Berkeley National Laboratory, 1 Cyclotron Road, Berkeley, CA 94720, USA\\
    $^{11}$Department of Physics, University of California Berkeley, 366 LeConte Hall MC 7300, Berkeley, CA 94720-7300, USA\\
	$^{12}$Research School of Astronomy and Astrophysics, Australian National University, Canberra, Australia\\
	$^{13}$Centre for Gravitational Astrophysics, College of Science, The Australian National University, ACT 2601, Australia\\
	$^{14}$School of Earth and Space Exploration, Arizona State University Tempe, AZ, USA\\
	$^{15}$Groupe d'Astronomie et Astrophysique, Universit\'e de Montreal, Montreal, Qu\'ebec, Canada \\
	$^{16}$Infrared Processing and Analysis Center, Pasadena, California, USA\\
	$^{17}$European Southern Observatory, Karl-Schwarzschild-Str. 2, 85748, Garching, Germany\\ 
}

\date{Accepted 2020 September 15. Received 2020 September 14; in original form 2020 August 31}

\pubyear{2020}

\begin{document}
\label{firstpage}
\pagerange{\pageref{firstpage}--\pageref{lastpage}}
\maketitle

\begin{abstract}
We calculate H$\alpha$-based star formation rates and determine the star formation rate-stellar mass relation for members of three SpARCS clusters at $z \sim 1.6$ and serendipitously identified field galaxies at similar redshifts to the clusters. We find similar star formation rates in cluster and field galaxies throughout our range of stellar masses.  The results are comparable to those seen in other clusters at similar redshifts, and consistent with our previous photometric evidence for little quenching activity in clusters.  One possible explanation for our results is that galaxies in our $z \sim 1.6$ clusters have been accreted too recently to show signs of environmental quenching.  It is also possible that the clusters are not yet dynamically mature enough to produce important environmental quenching effects shown to be important at low redshift, such as ram pressure stripping or harassment.
\end{abstract}




\section{Introduction}

The relationship between star formation and environment, its evolution with redshift, and the mechanisms for producing the relationship have been an ongoing matter of interest and debate.  The morphology-density relation \citep{dre80,pos05,vdw07,hol07,mei12,bas13} and star formation rate (SFR)-density relation \citep{ell01,kau04,pat11} observed at redshifts $z \lesssim 1$, and sometimes beyond, both indicate that typical galaxies are forming fewer stars and are more bulge-dominated in higher-density environments for a large portion of the age of the Universe.  Such findings have led to an emphasis on the search for causes and evidence of quenching of star formation via the heating or removal of the atomic and molecular gas supplies shown by \citet{ken98} to regulate star formation in galaxies.  Mechanisms such as harassment \citep{moo96} and ram pressure stripping \citep{gun72} are thought to operate primarily in clusters, although the latter may operate to a lesser degree in groups, especially on low-mass galaxies \citep[e.g.][]{fham16}. Strangulation \citep{bal00} and slow tidal interactions, including mergers of gas-rich galaxies, are thought to dominate in group environments, although the former also operates in clusters.  It is also commonly found that galaxies falling into clusters show signs of having already begun the quenching process in their previous group environment, a phenomenon known as pre-processing \citep{mcg09,wet15}. Given that high-mass galaxies also tend to be less star-forming independently of environment \citep{pen10}, stellar-mass-linked mechanisms such as bulge growth \citep{mar09,abr14} and AGN feedback \citep{cro06} are also thought to cause quenching.

The most common evidence of quenching mechanisms related to environment is a higher percentage of quenched galaxies in groups and clusters than in less dense environments \citep{bal04, blan05, coop06}. The enhancement of quenched fraction in clusters and groups appears to evolve with redshift. Some studies reported a reversal of the star formation rate-density relation for intermediate-density (group) environments at $z \sim 1$ \citep{elb07,coo08,sob11}, while others cast doubt on that claim \citep{zip14,lem19}.  However, studies of rich galaxy clusters at these redshifts \citep{muz12,nan13a,nan13b,pos05,vdw07,hol07,pat09,mei12} are consistent in finding that galaxies are more likely to be passively evolving than in the field just as at lower redshifts, with the possible exception of the lowest-mass galaxies in the surveys. 

Although the morphology-density and SFR-density relations in clusters have been found out to $z \sim 1.6$ in some cases \citep{bas13}, a scarcity of confirmed clusters and groups beyond $z \sim 1.5$ with passive or red fraction estimates leaves the situation unclear at earlier cosmic times.  A handful of studies exist at redshifts $z \geq 1.5$, most showing at least a moderate enhancement in passive or red fractions with respect to the field \citep{qua12, new14, cook16, lee17, str19}, though \citet{nan17} found only a mild enhancement. A number of observational and simulation-based studies \citep{tra10,tra17,bro13,alb16,hwa19} indicate that the SFR-density relation may start to break down in clusters at $z \geq 1.5$.

A more direct sign of quenching, equivalent to seeing it in action, would be environmental variation in another important scaling relation: the SFR-stellar mass relation \citep{noe07}, also known as the star formation main sequence or star formation mass sequence (SFMS).  This relation is found to exist with roughly the same slope for most of the known history of the Universe, but with a decline in median SFR with decreasing redshift for general field populations \citep{whi12,whi14,sch15}.  The higher median SFRs at higher redshifts allow for a variation on the quenching mechanism of strangulation, known as overconsumption \citep{mcg14}. In an overconsumption scenario, the environment removes external gas reservoirs via tidal stripping or gentle ram pressure (as in normal strangulation), and then the high SFR of the galaxy quickly exhausts whatever internal reservoirs exist in the galaxy's disk or bulge.

It is unclear how, and from which redshift, the environment of galaxies starts to affect the SFMS.  \citet{old20} has found some evidence that the slope of the SFMS may be slightly steeper due to suppressed SFRs for lower-mass star-forming galaxies in clusters since $z \sim 1.2$. A number of other studies find this relation to be apparently unaffected by environment at similar or higher redshifts \citep{muz12,koy13,tra10,tra17,erf16}.  This lack of direct evidence for environmental quenching is sometimes interpreted in terms of the delayed-then-rapid quenching model \citep{wet13} in which the effects of quenching on galaxy SFRs are not apparent until some time after the external gas reservoir is lost.  Some studies, such as \citet{old20} at $1 < z < 1.4$ and \citet{vul10} at lower redshifts, do find slightly lower SFRs in dense environments.  \citet{zei13} find such an effect as well, but limited to low-mass galaxies only.  Thus, there is still uncertainty as to whether the effects of quenching can be seen directly in galaxies that are in early stages of the process of quenching in high-density environments at intermediate redshifts.

The Spitzer Adaptation of the Red-Sequence Cluster Survey, or SpARCS \citep{muz09,wil09}, has provided one of the largest data sets for the exploration of the effects of very dense environments on galaxy evolution to date. Particularly valuable for the critical redshift range of $1.3 < z < 2$, when cosmic star formation begins to decline and mature clusters start to become common, is the SpARCS high-redshift cluster sample.  This subsample of confirmed SpARCS clusters consists of five systems, four (SpARCS-0224, SpARCS-0225, SpARCS-0330, and SpARCS-0335) in the southern hemisphere identified by the mid-IR identification of the rest-frame 1.6~$\mu$m stellar bump \citep{muz13b,nan16} and one (SpARCS-1049) in the northern hemisphere identified serendipitously by red sequence selection and first presented in \citet{web15}. Four of the five clusters (SpARCS-0224, SpARCS-0225, SpARCS-0330, and SpARCS-1049) were targeted for extensive ground- and space-based follow-up imaging and spectroscopy by the SpARCS team.  SpARCS-0335, lying at a significantly lower redshift than the other four ($z = 1.37$), was followed up later by \citet{bal17} and Balogh et al. (in press) as part of the Gemini Origins of Galaxies in Rich Early Environments (GOGREEN) survey.

The SpARCS high-redshift clusters, especially the three $z \sim 1.6$ southern clusters, are among the best studied at their redshifts. Their stellar mass functions and passive fractions from spectral energy distribution (SED) fitting indicate slight differences from contemporaneous field samples and large differences from lower-redshift SpARCS clusters, indicating rapid evolution in cluster galaxies \citep{nan16,nan17}.  \citet{lid12} used two of the three southern high-redshift clusters together with other SpARCS clusters to constrain the rate at which the stellar mass of brightest cluster galaxies (BCGs) increases, finding that it was best explained by dry mergers rather than last-minute star formation at redshifts between 1.6 and 0.2.  \citet{fol18} found a slightly shorter quenching timescale in the higher-redshift SpARCS clusters (including SpARCS-0335 and the $z \sim 1.6$ clusters) than the lower-redshift ones, consistent with quenching methods that depend on time spent in the cluster (e.g.~ram pressure). The ALMA CO emissions of star-forming galaxies in these clusters indicate possible enhanced gas fractions \citep{nob17,nob19}. Morphology with {\it Hubble Space Telescope} ({\it HST}) imaging \citep{del17} indicates that the merger rates in the three $z \sim 1.6$ SpARCS clusters plus SpARCS-1049 at $z \sim 1.7$ are similar to the field or slightly suppressed, making it unlikely that mergers are the main mechanism resulting in the quenching of these galaxies.  

The other two SpARCS high-redshift clusters have also appeared in several important studies independently of the $z \sim 1.6$ subsample. SpARCS-1049 is a very massive cluster for its redshift \citep{fin20} which has been the subject of extensive studies of its peculiar central starburst, suspected to be fed by a cooling flow or multiple wet mergers \citep{web15,web17,bon17,tru19,hl20}. SpARCS-0335, as part of the GOGREEN survey, has been included in almost all of the GOGREEN early science papers including \citet{old20}, \citet{vdb20}, Chan et al.~(in press), and Webb et al.~(in press).

In this paper, we estimate the H$\alpha$ SFRs and study the SFMS of cluster and serendipitous field galaxies from the three $z \sim 1.6$ clusters of the SpARCS high-redshift sample.  Our goal is to constrain the nature of the quenching processes acting in star-forming cluster galaxies: substantial drops in SFRs in clusters would support slow quenching, while field-like SFRs would support fast quenching or a delayed-then-rapid mechanism. In Section 2, we describe the data and our data reduction techniques.  In Section 3, we describe our techniques for data analysis.  In Section 4, we present our results, and in Section 5 we discuss their importance.  Throughout the paper, we use a flat cosmology with $\Omega_{\Lambda} = 0.7$, $\Omega_{M} = 0.3$, and $h = 0.7$.

\section{Data and analysis}

\subsection{Multi-Band Photometry}
Our three southern high-redshift clusters, SpARCS-0224, SpARCS-0225, and SpARCS-0330, all have multi-band photometry ranging from the $u$-band (or $g$-band in SpARCS-0225) to the IRAC 8~$\mu$m band.  Most of the photometry was described in \citet{nan16}.  The ground-based photometry, which has an angular area of 8.4$\times$8.4 arcminutes, was used to select the galaxies for spectroscopic follow-up described in this paper.  Within this region, above the passive-galaxy photometric completeness limits of our data as described in \citet{nan16} (10$^{10.4}$ M$_{\odot}$), each of our three clusters had 3-4 times as many galaxies at photometric or spectroscopic redshifts consistent with cluster membership (90-108 each) as the average of the UltraVISTA data release 1 \citep{muz13a} field expected in the same angular area (26-31 per individual cluster).

To our original ground-based photometry we have recently added photometry from the analysis of {\it Hubble Space Telescope} ({\it HST}) images in $F105W$ ($Y$), $F140W$ ($J$) and $F160W$ ($H$) \citep{del17}.  The $F105W$ and $F140W$ images, for the clusters SpARCS-0224 and SpARCS-0330, come from the See Change survey (Hayden et al., submitted) and cover about 2.8$\times$2.8 arcminutes centered on an area close to the clusters' brightest galaxies.  All three clusters have $F160W$ imaging obtained by the SpARCS team, with similar area coverage.  Approximately 50\% of the confirmed cluster members used in our analysis are found within the coverage of the $HST$ imaging.

We used the $HST$ images for two purposes: (a) adding more aperture photometry in order to refine photometric redshifts, stellar masses, and dust extinction estimates and (b) estimating, or calibrating uncertainty margins in estimates for galaxies without $HST$ coverage, the total fluxes in $H$ in order to perform the final flux calibrations on the spectroscopy. The first usage is discussed in this section, and the second usage is discussed in Section 2.3.

In order to use the aperture photometry to refine SED-based parameters, we matched the $HST$ images to the point spread function (PSF) of the lowest-resolution ground-based image and performed the aperture photometry and random error estimates using the same technique described in Section 3.1 of \citet{nan16}. We then re-ran EAZY \citep{bra08} and FAST \citep{kri09} on the new set of aperture photometry to update the SED-derived parameters  including stellar masses, dust extinctions, and rest-frame colors for galaxies in the $HST$ images.  We also obtained new SED-derived parameters for galaxies with new redshifts found since 2016 in the reanalysis of the spectra (Section 2.2) and in the ALMA CO studies of \citet{nob17,nob19} and van Kampen et al.~(in preparation).

The only significant change in the values of SED-derived parameters for most galaxies with $HST$ coverage was in the rest-frame $UVJ$ colors.  These colors were typically offset by up to 0.1 mag bluer in $U-V$ and 0.1 mag redder in $V-J$ compared to the old photometry, primarily due to the better-constrained flux in the observed $H$ band (between rest-frame $V$ and $R$ at $z = 1.6$) resulting in the EAZY code estimating lower fluxes in this wavelength range.  The difference was less dramatic for SpARCS-0330 and SpARCS-0224, which already had ground-based $J$-band data, than for SpARCS-0225, which previously had no photometry between $Y$ and $K$.  When we analyzed the $UVJ$ color distribution for the objects with $HST$ imaging, we found that the loci of passive and star-forming galaxy concentrations were consistent with needing to shift the passive and star-forming cutoff lines according to the rest-frame color changes in order to separate the populations.  In other words, the old $UVJ$ cutoff criteria based on \citet{whi11} with the old photometry appear to be accurate in separating passive from star-forming galaxies, even if the new colors are more accurate for these galaxies.

\subsection{Spectroscopic Data and Reduction}

Our spectroscopic data for this study were obtained with the Multi-Object Spectrograph For InfraRed Exploration (MOSFIRE) \citep{mcl10,mcl12} on the Keck I telescope on Mauna Kea, Hawaii, between November 2012 and November 2014 (PI G. Wilson).  These observations were planned as part of a follow-up campaign on suspected or confirmed z $>$ 1.5 clusters. The southern clusters in this study were previously confirmed with European Southern Observatory (ESO) Very Large Telescope (VLT) Focal Reducer/low dispersion Spectrograph 2 (FORS2) spectra \citep{muz13b,nan16}.  We obtained a total of 588 spectra of 561 science objects on 8 masks for SpARCS-0330, 6 masks for SpARCS-0224, and 2 masks for SpARCS-0225.  Redshifts for these clusters derived from emission lines identified in the 2-D spectra were presented in \citet{nan16}.  The flux calibration was recently performed for this paper and is described in the next subsection.  A total of 370 of our science objects have redshift estimates; the remaining objects did not have spectra with sufficient signal-to-noise (S/N) to estimate the redshifts, or had no emission lines within the $H$ band (e.g.~science stars, objects with z $<$ 1.2.)  See Appendix for a summary of the observations.

\begin{figure*}
      \begin{subfigure}
      \centering
      \includegraphics[width=\columnwidth]{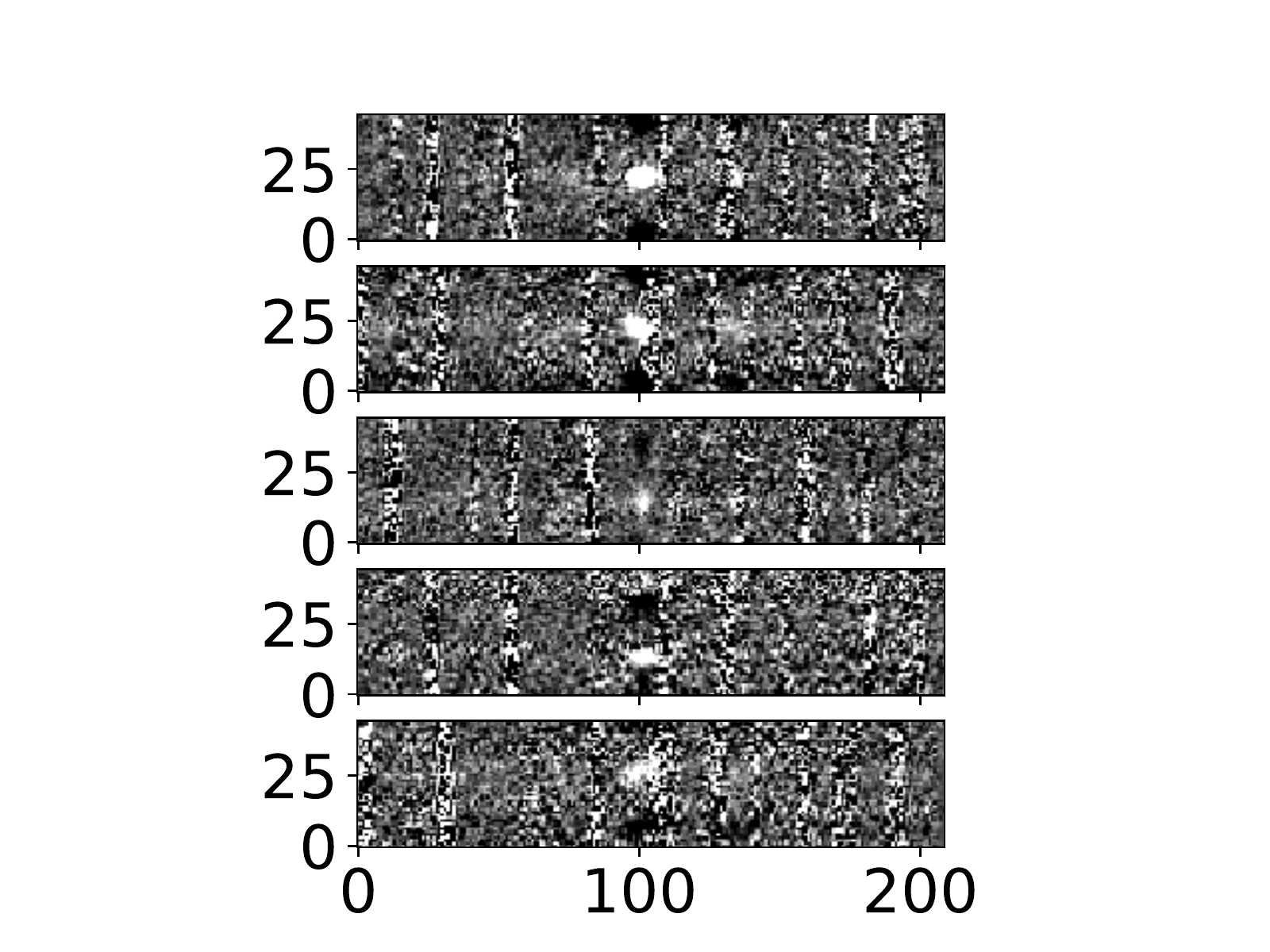}
      \end{subfigure}
      \begin{subfigure}
      \centering
      \includegraphics[width=\columnwidth]{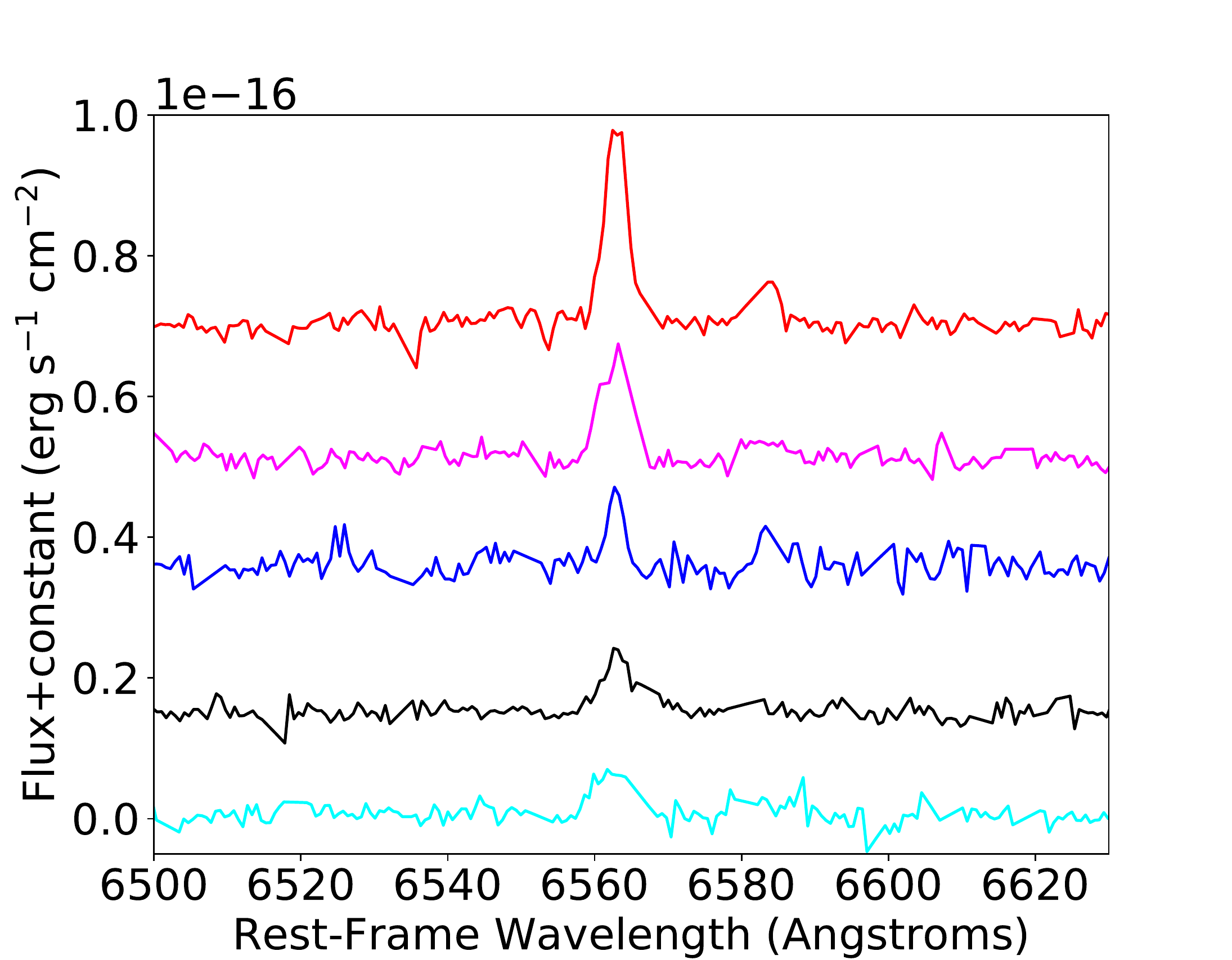}
      \end{subfigure}
      \caption{Left: Cutouts of two-dimensional reduced spectra of five cluster members from SpARCS0330, with clear H$\alpha$ emission lines of different intensity and shape. Right: One-dimensional extractions, shown with rest-frame wavelengths and on a common flux scale, of the same five cluster members shown in the same order as the left panel.}
      \label{fig:spec}
\end{figure*}

Our spectra were reduced with a slightly modified version of the public MOSFIRE data reduction pipeline,\footnote{https://keck-datareductionpipelines.github.io/MosfireDRP} which summed the exposures of each mask, extracted and rectified the 2-D slit spectra, calibrated the wavelength scale using telluric emission lines, subtracted these telluric lines, and performed 1-D spectral extraction.  We modified the pipeline to overcome problems with the reduction of standard star data, and to allow a customized extraction of the 1-D spectra to screen out as many telluric lines as possible while preserving genuine emission lines.  Our customization of the 1-D extraction involved performing an ``optimal'' extraction \citep{hor86} after the removal of any regions where the noise in the 2-D rectified spectrum summed across the column was more than 2.5 $\sigma$ brighter than the median for the whole spectrum (all rows and columns).  This requirement to sum across the column prevented genuine emission lines, which could sometimes be very bright but which were not extended across the entire column, from being screened out as telluric lines.  After each 1-D extraction, the spectrum was visually compared to the 2-D spectrum in order to identify emission lines in both.  No bright lines were absent in the 1-D spectrum that were present in the 2-D spectrum (indicating that our method for screening sky lines did not eliminate bright emission lines), although some faint lines in the 2-D spectrum failed to rise visibly above the noise in the 1-D spectrum.  Figure~\ref{fig:spec} shows extracted 2-D spectra of five cluster members in SpARCS-0330 with different H$\alpha$ line strengths in a region near the H$\alpha$ line, along with the extracted, rest-frame-corrected, and flux-calibrated 1-D spectra in the same region with a constant added to the fluxes of the top four spectra for easy comparison of line strengths.

We performed 1-D extraction of our A0V standard stars using a standard rather than ``optimal'' extraction algorithm, since the standard extraction produced a slightly higher S/N spectrum for these objects.  We used ESO's {\sc{molecfit}} code \citep{sme15,kau15} to remove the telluric absorption features from the standard star spectra. We then fit a polynomial to the ratio of the standard star spectrum and a generic A0V spectrum from the \citet{pic98} library with absorption lines removed to produce sensitivity curves.  The polynomial fit excluded regions at very short (below 1.5~$\mu$m) and long (above 1.75~$\mu$m) wavelengths in which the spectrum lacked adequate data or S/N, so as to provide a smoother and more precise fit to the rest of the spectrum.

We flux-calibrated our 1-D science spectra in three steps.  First, we used {\sc{molecfit}} on each mask's science star to remove the telluric features from all science spectra in the mask.  Second, we applied our sensitivity curve derived from the A0V standard stars to all the science spectra in each mask.  Finally, we scaled the flux according to the $HST$ $H$-band or linearly interpolated ground-based aperture-corrected photometry (between $J$ or $Y$ and $K$ at the central wavelength of $H$) for all objects within our empirical limit for visual continuum detection in 1-D spectra.  This empirical limit corresponded to a total uncalibrated flux in the extracted spectrum at least 7\% of that in the RMS spectrum, which includes noise from all telluric lines and from low-sensitivity regions of the spectrum.  If the global S/N fell below this limit, or we did not have reliable near-IR photometry for the object, the science star was used to determine the flux scaling for the object.  In rare cases of masks with two science stars, we used the average of the two science stars for the flux scaling for objects with no detectable continuum or no reliable near-IR photometry.

Comparing interpolated, aperture-corrected photometry to the total $F160W$ photometry for objects above our $K$-band completeness limits, we estimate a typical calibration error (mean absolute offset) of 9\% for SpARCS-0330 and SpARCS-0224 and 20\% for SpARCS-0225.  The latter is most likely higher due to the absence of $J$-band photometry to constrain the interpolation.  This error margin was added in quadrature, per pixel, to the original (flux-calibrated) RMS spectrum for all objects bright enough to be calibrated via their own photometry.

In order to estimate the total error for objects calibrated via the science stars rather than their own photometry, we compared the fluxes derived from an object's own photometry (meeting continuum and aperture photometry requirements) to those derived from the science star as was done for objects with poorer data. Across all masks, the median difference between the two flux calibrations was 38\%.  This result is roughly consistent with a quadrature sum of the photometry-related uncertainty of 9\% or 20\% described above and a 35\% uncertainty related to seeing, mask position, and telluric corrections, derived from comparing the uncalibrated fluxes of objects observed in two different masks which met our criteria for calibration via their own photometry as described above. The 38\% uncertainty margin was therefore added in quadrature per pixel to the RMS spectra calibrated with the science star.


\subsection{Analysis}
Our cluster members are selected to have $\delta_{z\_spec}$ within 0.015 of their respective cluster BCG, as in our previous papers \citep{nan16,nan17}.  This is a generous cut unlikely to exclude any genuine members but at risk of including some interlopers.  We found no major changes to our results using the less generous cut of 0.010.  Non-member galaxies in the cluster field for further analysis were restricted to redshifts above 1.5 in order to minimize the effects of star formation rate evolution on our results, and to screen out the $z = 1.43$ group in the foreground of SpARCS-0225 previously identified with FORS2 spectra \citep{nan16}. No cluster-centric distance limit was used to distinguish members from nonmembers, so as to allow the inclusion of galaxies in smaller subgroups or filaments that will likely merge with the cluster core by $z = 0$. The maximum redshift for analysis for nonmembers is $z \sim 1.7$, the highest redshift at which H$\alpha$ is visible in the $H$ band.

The H$\alpha$ flux and its uncertainty were estimated via a Monte Carlo simulation of 1000 single Gaussian fits to the flux-calibrated 1-D spectrum varied within its flux-calibrated RMS spectrum (including systematic uncertainty measurements discussed above) at the estimated position of the H$\alpha$ line based on the individual galaxy's spectroscopic redshift.  The median of these fits was taken as the H$\alpha$ flux, and the 68\% confidence levels above and below this median flux estimate were taken as the total flux uncertainty. The detection S/N for the H$\alpha$ line was estimated by performing the same fits on the uncalibrated spectrum, using the fitted peak value minus the fitted continuum value (that is, the portion of the peak attributable to the line only) as the signal and the 1$\sigma$ (68\%) uncertainties in the fitted continuum as the noise, and dividing the former by the latter.  

We converted the H$\alpha$ fluxes to luminosities using the luminosity distance to the object calculated via its redshift within our cosmology of choice.  We corrected the H$\alpha$ flux for the estimated internal reddening based on SED fitting to our multi-wavelength photometry, updated in this paper to include photometry from the $HST$ imaging described in \citet{del17}.  The average SED extinction value was ${\rm A}_{V} \sim 1.09$, with a scatter of 0.8.   We then multiply by the stellar-to-gas extinction correction factor of 1.86 from \citet{pri14}, to account for the line emission usually coming from the dustiest parts of the galaxy.  We then converted this luminosity to a SFR using the conversion factor of \citet{ken98} corrected to a \citet{cha03} initial mass function (IMF), as used in \citet{twi12}.

Our H$\alpha$ derived SFRs are comparable to those of \citet{nob17,nob19}, based on mid-IR fluxes in massive galaxies.  For the seven galaxies that have viable SFRs both in this work and \citet{nob17,nob19} the median difference is very small (0.02 dex, with the H$\alpha$ SFRs higher).  If we do not apply the SED-to-gas reddening correction from \citet{pri14}, our SFRs are about 0.4 dex lower than the mid-IR-based values, comparable with those of \citet{tra17}, who did not apply a stellar-to-gas extinction correction factor to their SFRs.

\begin{figure}
    \centering
    \includegraphics[width=\columnwidth]{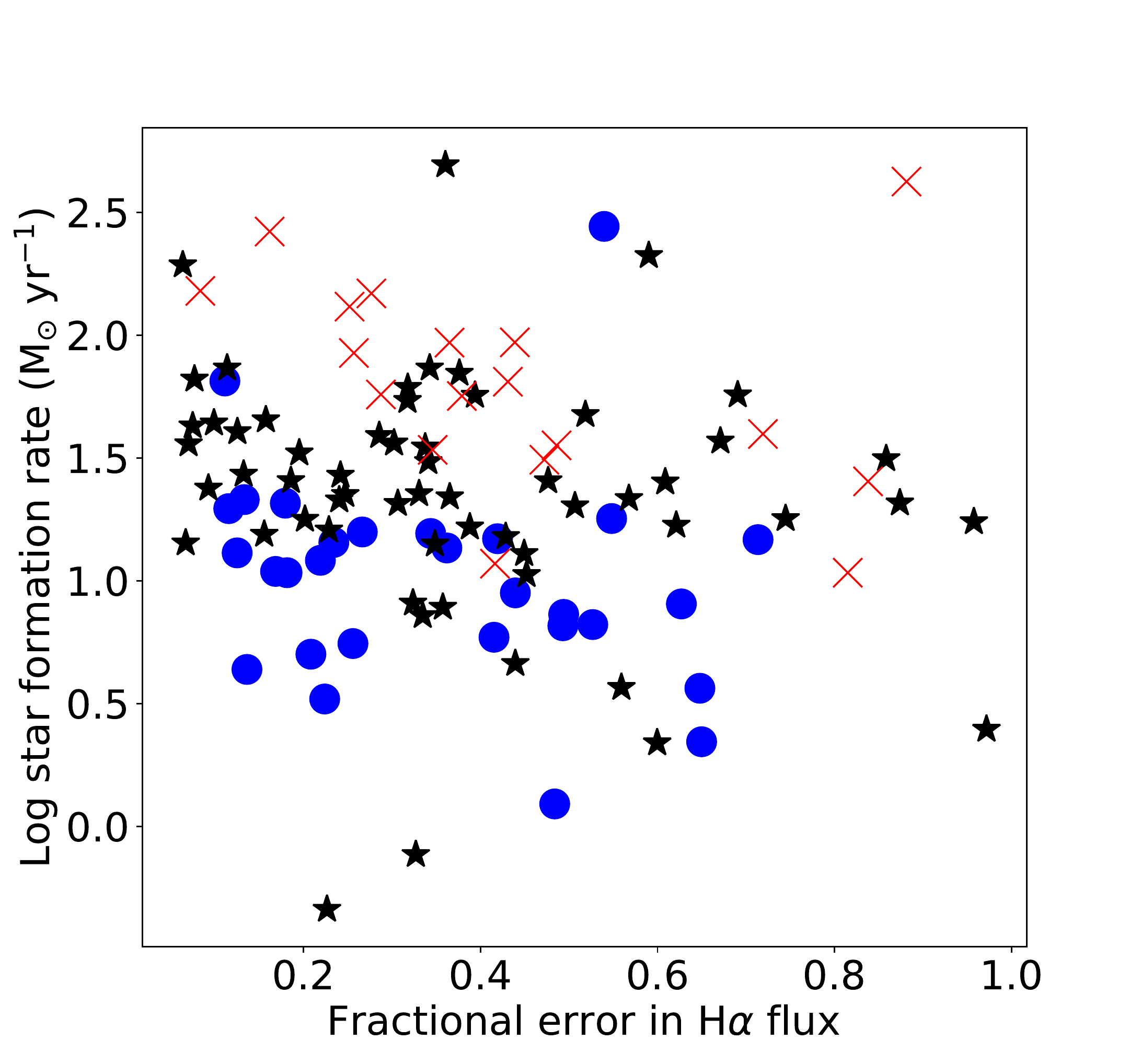}
    \caption{Individual log SFRs versus fractional uncertainty in the H$\alpha$ flux (from line fitting) for all galaxies, cluster members and non-members alike, selected for analysis. Blue circles represent low-mass galaxies with 9 $<$ log (M$_*$/M$_{\odot}$) $<$ 9.7, black stars are intermediate-mass galaxies with 9.7 $<$ log (M$_*$/M$_{\odot}$) $<$ 10.4, and red crosses are high-mass galaxies with 10.4 $<$ log (M$_*$/M$_{\odot}$) $<$ 11.0. Most galaxies within the typical range of fractional uncertainty have SFRs of at least 6-10 M$_{\odot}$ yr$^{-1}$ depending on their mass range, with a handful of objects with lower fractional uncertainty having lower SFRs.}
    \label{fig:logsfr_vs_fracerr}
\end{figure}

All further analysis as described below was performed only on objects (a) with fractional error $\leq$ 1 or H$\alpha$ line-fitting S/N $\geq$ 1 (from Monte Carlo trials of Gaussian fits to calibrated spectra, distinct from detection S/N), (b) whose detection S/N from the uncalibrated spectrum was at least 3, and (c) whose stellar mass was between $10^9$ and $10^{11}$ M$_{\odot}$. Figure \ref{fig:logsfr_vs_fracerr} shows the distribution of SFRs as a function of fractional uncertainty $\sigma_{H\alpha}/H\alpha$ for selected objects in the clusters and field. Our fractional error limit usually corresponds to a minimum H$\alpha$ flux of $10^{-16.5}$ erg s$^{-1}$ cm$^{-2}$ or a minimum SFR of about 6 M$_{\odot}$ yr$^{-1}$ for typical low- (blue) and intermediate-mass (black)  galaxies. Our high-mass star-forming galaxies (red) have a minimum SFR of about 10 M$_{\odot}$ yr$^{-1}$.

The stellar mass cutoffs make the bins smaller and more comparable between the two environments in their median stellar mass, preventing unusual galaxies like low-mass starbursts or high-mass brightest cluster galaxies from dominating the results in a bin, while the other cutoffs screen out spurious H$\alpha$ detections.  The three requirements resulted in 47/97 members and 59/131 nonmembers with redshifts above 1.5 in the final sample for analysis (all flux calibrated with their own photometry).  The flux uncertainty cuts dominated the selection of objects, having substantial but not exact overlap with the detection S/N cuts, while the stellar mass cut eliminated a much smaller portion of galaxies. No obvious Type 1 AGN were found in the sample after applying the other selection processes.  A $UVJ$ color cut was made unnecessary by the other cuts, since no $UVJ$ passive galaxies via the color selection based on \citet{whi11} as used in \citet{vdb13} within our stellar mass limits made the flux uncertainty cut. 

Due to the large uncertainties and high scatter in individual SFRs, we binned our data according to stellar mass.  Three bins were defined for both cluster and field galaxies: log($M_{*}/{\rm M}_{\odot}$) = 9--9.7, log($M_{*}/{\rm M}_{\odot}$) = 9.7--10.4 (the most populous bin given the combination of the inherent stellar mass function and spectroscopic detection limits), and log($M_{*}/{\rm M}_{\odot}$) = 10.4--11.0. We performed a Monte Carlo simulation re-selecting our objects for each bin 100 times after varying the SED-derived stellar mass within its uncertainties, calculated the median SFR in each re-selection, and took the median of these trials as the SFR for the bin and the 68\% confidence levels as the uncertainties.  The Monte Carlo median SFR uncertainties in the bins ranged from 0.08 to 0.19 dex (20 to 55 percent), being smallest in the most populous bin of intermediate stellar mass.  Using means instead of medians had no effect on our results other than to increase the uncertainty margin.

Because of the small but substantial difference in median redshift of the field and cluster samples (1.57 and 1.62 respectively), we used the \citet{sch15} formula for the evolution of the SFMS with redshift to estimate a correction for the field SFRs.  The correction was determined to be the difference between the \citet{sch15} typical SFR for a galaxy at the median redshift of the clusters with the same stellar mass as the field galaxy or bin in question and that corresponding to the genuine redshift (or median bin redshift) of the field galaxy. We estimated this correction separately for the bins described above and for the individual SFRs used for the SFMS, and added it to the logarithm of the respective field galaxy SFR.  Typical corrections were on the order of 0.01-0.02 dex, substantially lower than our error margins, but had a slight effect on the apparent degree of (in)significance of any differences between the cluster and field.

\section{Results}

\begin{figure*}
	\includegraphics[width=17cm]{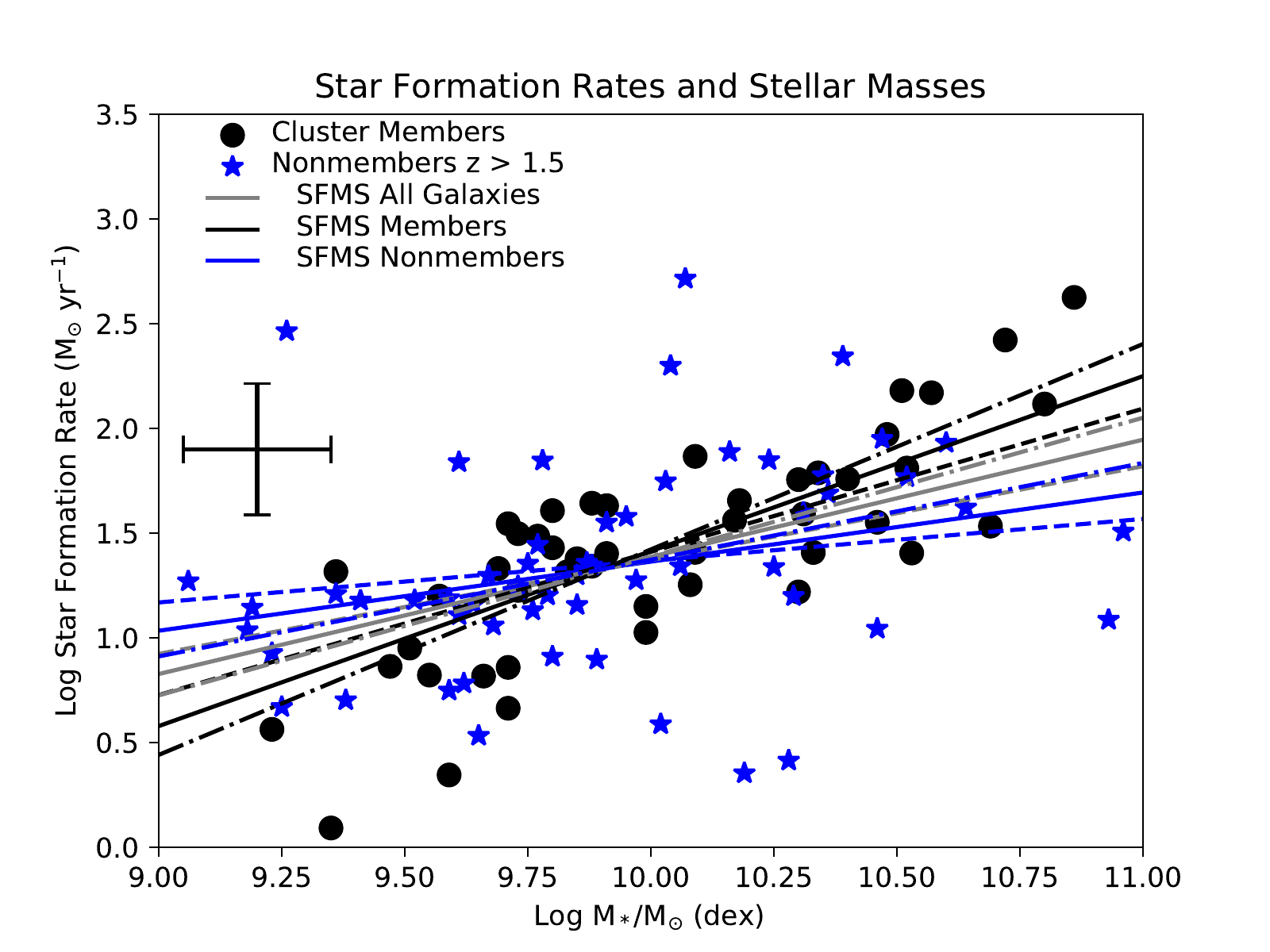}
	\caption{Individual SFRs versus stellar mass for star-forming galaxies meeting our selection criteria in the three clusters (black circles) and the field (blue stars).  The values for field galaxies are corrected for estimated evolution of the SFMS with redshift.  Also shown are least-squares, unweighted linear fits to the collective (gray), field (blue), and cluster (black) distributions, and typical individual uncertainties (black error bars below the legend text).  The 68\% uncertainty margins to the linear fits are shown as dashed (lower slope/upper intercept) and dash-dot (upper slope/lower intercept) lines for the combined sample (gray), field (blue), and clusters (black).} 
	\label{fig:sfrs}
\end{figure*}
\begin{figure*}
	\includegraphics[width=17cm]{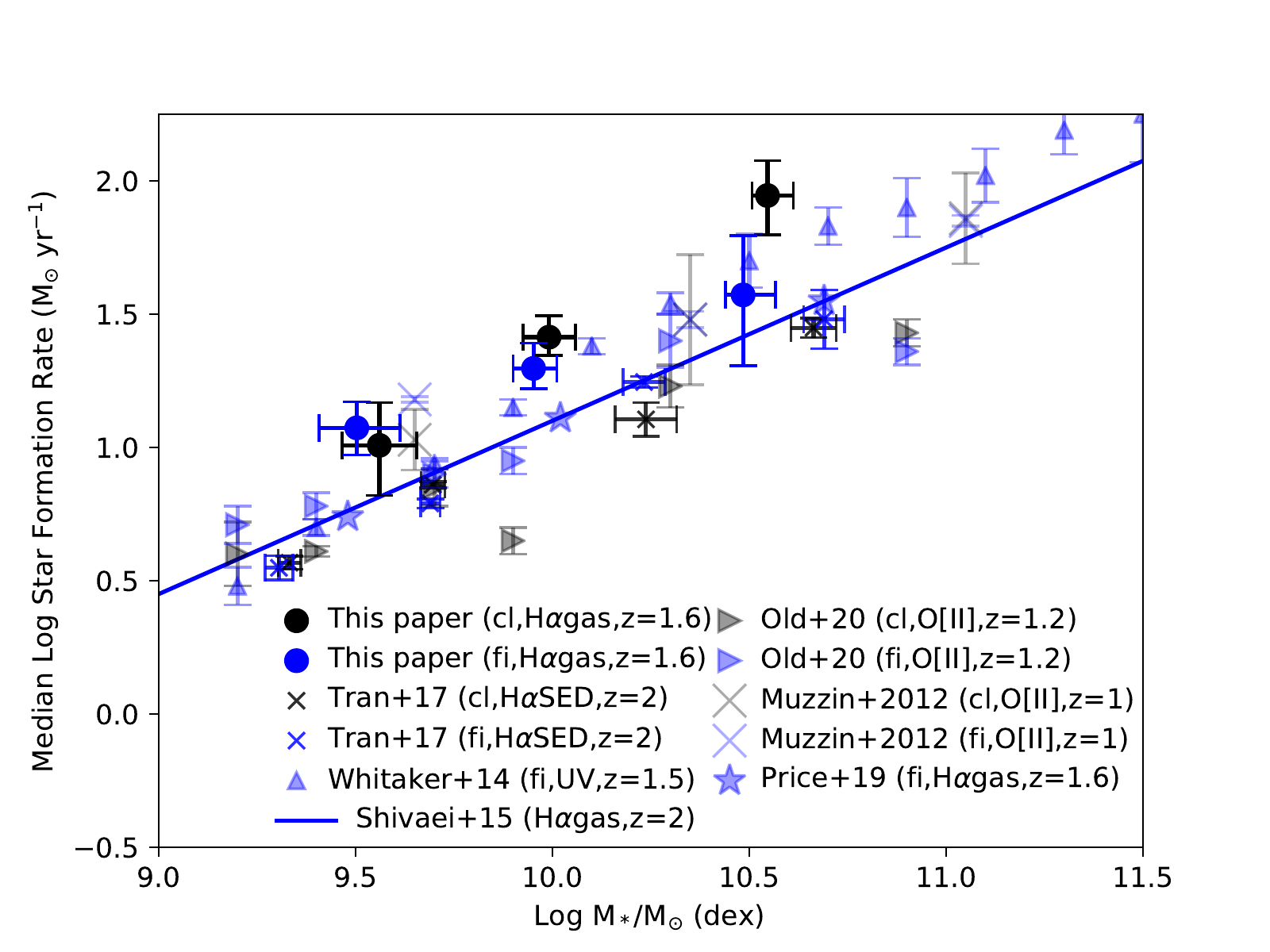}
	\caption{Comparison of our binned results (members as black circles and nonmembers corrected for star formation rate evolution with redshift as blue circles) with selected values from the literature, designated in the legend below the plot.  Our SFRs are similar to many other works using optical emission lines, with some works finding substantial differences between cluster members and field galaxies and others finding little to no difference like our study.  Some of the differences found in other studies are smaller than our uncertainties at the corresponding stellar masses.}
	\label{fig:compare}
\end{figure*}

\begin{figure}
	\includegraphics[width=\columnwidth]{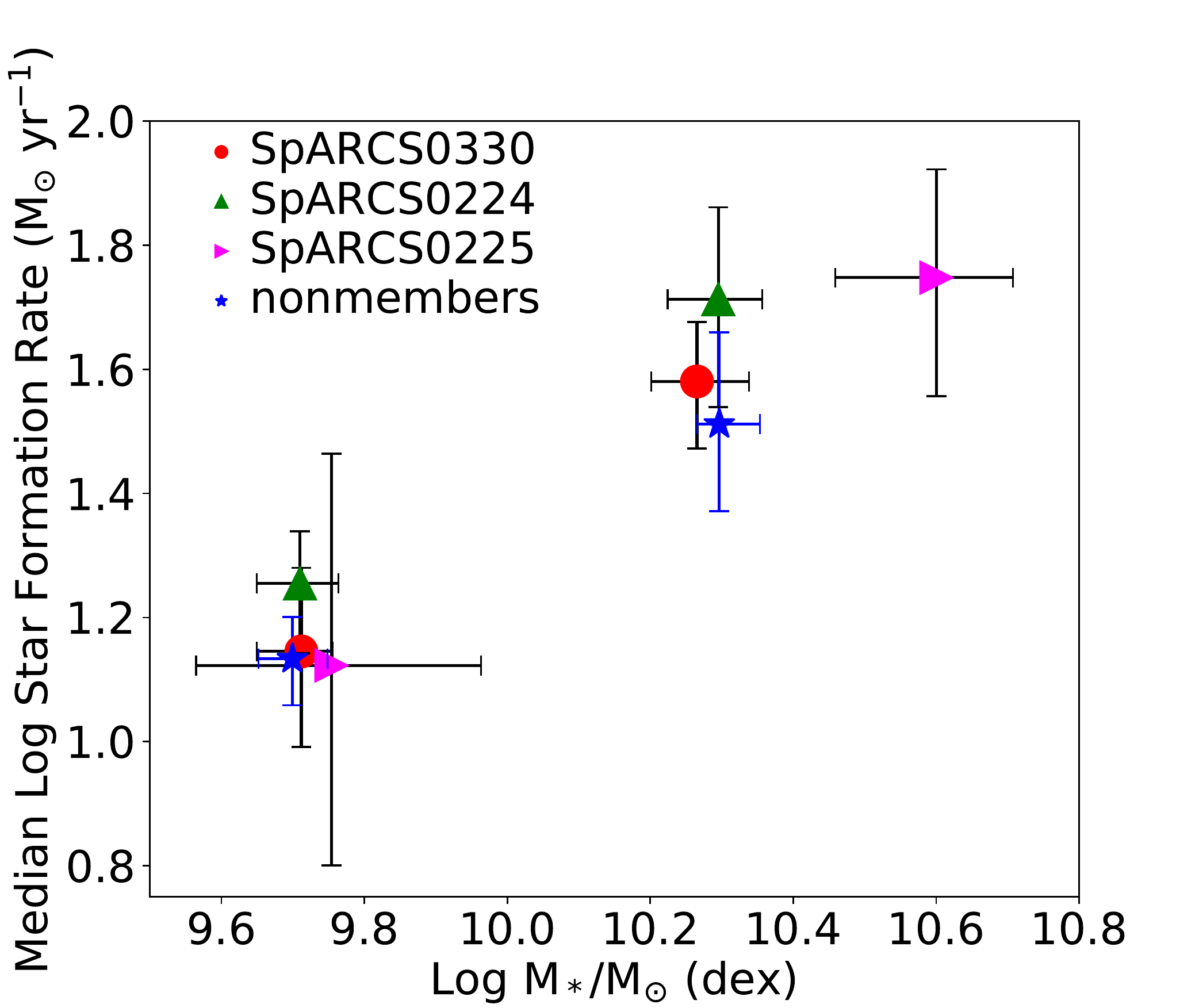}
	\caption{Comparison of the star formation rate-stellar mass relations of the individual clusters within our sample and the nonmembers with redshifts above 1.5, using two stellar mass bins for each cluster. (The median stellar mass of SpARCS-0225 high-mass star-forming galaxies is higher than in the other clusters.)  The field values are corrected for evolution in the star formation main sequence with redshift.}
	\label{fig:separatesfms}
\end{figure}

Figure \ref{fig:sfrs} shows the individual SFRs for cluster members (black circles) and non-members with \citet{sch15} corrections to the SFR (blue stars).  All objects shown are star-forming galaxies with line flux uncertainty ($\sigma_{H\alpha}/H\alpha) \leq 1$ and detection S/N of at least 3, or at least about 6 M$_{\odot}$ yr$^{-1}$ for most spectra (see Figure \ref{fig:logsfr_vs_fracerr}). The figure also shows unweighted least-squares linear fits (solid lines) with upper slope/lower intercept (dash-dot lines) and lower slope/upper intercept (dashed lines) uncertainty margins to the full distribution (gray), clusters only (black), and field only (blue), preferred over the weighted fits so as not to over-count exceptionally bright and high S/N spectra.  Uncertainty-weighted fits mostly ignored the more common medium and low star formation rates or, if forced to pass through the median, yielded very large uncertainties in the slope and intercept.  The average slopes and intercepts for the SFMSs and their 68\% error margins were estimated with 1000 instances of random bootstrap resampling of the original cluster and field galaxy samples, reapplying with each trial the uncertainty, detection S/N, and stellar mass cuts. The linear fits to the bootstrap samples for the full sample, clusters, and field were each forced through the median stellar mass ($9.9 \pm 0.5$) and median star formation rate ($1.3 \pm 0.4$) of the full bootstrap sample.  This was achieved by adding the median mass and median star formation rate as an artificial data point and setting its weight to 100 while all other weights were set to 1.  This step was meant to help control for differences at the extremes of the sample ruled by small number statistics, but performing pure unweighted fits without forcing them through the full sample medians produced virtually identical values and uncertainties.

We found a slope of $0.83 \pm 0.15$ and intercept of $-6.9 \pm 1.5$ for the clusters, and a slope of $0.33 \pm 0.14$ with an intercept of $-2.0 \pm 1.4$ for the field.  For the cluster and field points together, the collective slope was $0.56 \pm 0.11$ and the collective intercept was $-4.2 \pm 1.1$.  The cluster and field main sequences appear to differ in this analysis ($\sim 3.3 \sigma$), but this effect reduces to insignificance in the binned analysis below. Redoing the fits above with the lowest and highest 5\% of the stellar masses removed from the sample changed the cluster slope and intercept to $0.77 \pm 0.16$ and $-6.2 \pm 1.6$, and the field slope and intercept to $-0.38 \pm 0.19$ and $-2.5 \pm 1.9.$
 This represents a small numerical difference but, in combination with the larger uncertainties, it reduces the significance to a marginal 2$\sigma$. In Table \ref{tab:fracs}, we show the mean percentages of galaxies above and below the collective SFMS in our bootstrap trials, with 68\% random bootstrap resampling uncertainties. The percentage above and below the main sequence is essentially identical for low-mass galaxies ($<10^{10}$ M$_{\odot}$) in both types of environments and within $\sim 1.5 \sigma$ for high-mass galaxies.

We compare our SFMS with three mass bins in Table \ref{tab:sfms} to select observational data from the literature in Figure \ref{fig:compare}. For the non-member bins, we estimate a small correction for evolution with redshift according to the observational results of \citet{sch15}.  Our results are consistent (subtracting the 0.4 dex for SED-based vs. gas-based reddening corrections) with \citet{tra17}, who, like \citet{muz12}, find star-forming galaxies in clusters and field to be alike in emission-line-based SFRs. Our results are comparable to the \citet{whi14} curve for high stellar masses and slightly above the field galaxy results of \citet{shi15} and \citet{pri19} at similar redshifts, within 1-2$\sigma$ of the latter.  As might be expected, most of the results shown here, including ours, are comparable to the UV-based results of \citet{sch15}, generally lying between their $1.2<z<1.8$ and $1.8<z<2.5$ curves.

While \citet{old20} report a small difference between cluster and field [O{\scriptsize II}]-based SFRs in certain stellar mass bins, we find overall similarity with a hint ($\sim$1.4$\sigma$) of elevation of the median SFR at stellar masses above 10$^{10.4}$ M$_{\odot}$.  If we had twice as many galaxies in each bin with the same distribution of SFRs and stellar masses, the slight elevation in the high-mass bin would achieve marginal significance (2$\sigma$) comparable to \citet{old20} but in the opposite direction.  We attribute this primarily to statistical fluctuations due to the small number of high-mass galaxies (11 cluster galaxies and 7 field galaxies) with SFRs which meet our criteria.  Our result can still be considered consistent with \citet{old20} since the differences found in their study are on the level of the uncertainty margins in our own study. Also, our SFRs are similar to theirs in spite of their use of [O{\scriptsize II}], which suffers from greater dust extinction and more difficulty discerning contributions from AGN activity than H$\alpha$.

We also checked how the SFMSs compare individually among the three clusters.  As shown in Figure \ref{fig:separatesfms}, the clusters are all consistent with each other within approximately 1$\sigma$, although small sample sizes per cluster result in relatively large uncertainties.  SpARCS-0225 appears set apart from the others because its high-stellar-mass bin consists of 6 high-mass star-forming galaxies, of higher stellar mass on average than the smaller numbers of high-mass galaxies in the other two clusters (2 in SpARCS-0224 and 3 in SpARCS-0330).  Given the small numbers of galaxies involved, we consider this to most likely be a matter of random variation. The high-mass star-forming galaxies of SpARCS-0225 have about the same SFRs as those of SpARCS-0224, and therefore this cluster does not appear to be anomalous.

\begin{table}
	\caption{Fractions Below Collective Star Forming Main Sequence}
	\label{tab:fracs}
	\centering
	\begin{tabular}{lccc}
		\hline\hline
		Environment & Mass Range & Fraction below SFMS  \\  
		\hline
		Field & Low & 0.49$\pm$0.10 \\
		Field & High & 0.51$\pm$0.10 \\
		Clusters & Low & 0.46$\pm$0.09 \\
		Clusters  &High & 0.37$\pm$0.10 \\
		
		\hline  
	\end{tabular}
\end{table}

\begin{table*}
	\caption{Binned Median Star Formation Rates}
	\label{tab:sfms}
	\centering
	\begin{tabular}{lccc}
		\hline\hline
	Environment & Log(mass) & Log(sfr) & N  \\ 
	 	& M$_{\odot}$ & M$_{\odot}$/yr & \\ 
		\hline

		Field & 9.50 $\pm$ 0.10 & 1.07 $\pm$ 0.11  & 19 \\
		Field & 9.95 $\pm$ 0.06& 1.30 $\pm$ 0.09 & 33 \\
		Field & 10.48 $\pm$ 0.06 & 1.57 $\pm$ 0.24 & 7 \\
		Clusters & 9.56 $\pm$ 0.09 & 1.01 $\pm$ 0.18 & 10\\
		Clusters & 9.99 $\pm$ 0.07  & 1.41 $\pm$ 0.08 & 26 \\
		Clusters & 10.55 $\pm$ 0.05 & 1.95 $\pm$ 0.14 & 11 \\

		\hline  
	\end{tabular}
\end{table*}

\section{Discussion}
Our results indicate an overall similarity between clusters and the field in terms of SFRs of star-forming galaxies.  In \citet{nan17} we found passive fractions in these same clusters only slightly elevated compared to the field. This suggests that environmental quenching may not be fully operating in our $z \sim 1.6$ galaxy clusters.

We tentatively interpret our results on the basis that there has been little time for galaxy clusters to differentiate themselves from the field since the formation of the clusters. Perhaps most of the galaxies fell in the clusters too recently to show environmental quenching (i.e.~a full delay time has not yet passed), and/or the clusters are not yet virialized and thus have fewer means of quenching galaxies than more mature clusters. Our results are very similar to those of \citet{tra17}, which found no difference in the SFMS between their cluster galaxies and the field at about the same redshift as our own clusters.  \citet{muz12}, working at $z \sim 1$, also found no difference between the typical SFRs of cluster and field star-forming galaxies.  They did, however, find a substantial difference in the passive fractions and thus the collective SFRs of passive plus star-forming galaxies between cluster and field environments, similar to \citet{cha19} and \citet{vdb20} and unlike \citet{nan16,nan17}. Therefore our results are within expectations based on existing literature.

The combination of star formation rates similar to the field with passive fractions that are only modestly higher than the field \citep{nan17} suggests a genuine near-nullification of the SFR-density relation at $z \sim 1.6$ up to the low-mass cluster level. At lower redshifts, some low-mass galaxies show enhanced SFRs in moderate overdensities such as cluster outskirts and groups \citep{sob11}.  At slightly higher redshifts, in $z \sim 2$ protoclusters, \citet{tad19} found that low- and intermediate-mass protocluster galaxies had higher gas fractions and lower depletion times than contemporaneous field galaxies.  In the same three clusters studied in our paper, \citet{nob17,nob19} found evidence for enhanced gas fractions that could be a signature of previous protocluster or group enhancement of gas accretion. Given these types of findings, a lack of difference between the SFMS among cluster members and field galaxies is well within expectations.

\section{Conclusions}
We analyzed H$\alpha$-based SFRs as a function of stellar masses derived from SED fitting of multi-band photometry for the members of three $z \sim 1.6$ galaxy clusters from SpARCS and serendipitously identified non-members from the same data set at $1.5 \leq z \leq 1.7$.  We found that, consistent with some studies at similar or lower redshifts \citep{tra17,muz12} and subtly different from others at lower redshifts \citep{old20}, the SFRs of roughly contemporaneous, similar-mass cluster and field star-forming galaxies were statistically similar to one another. Our results combined with those of \citet{nan17} suggest there may be a nullification of the SFR-density relation up to densities corresponding to lower-mass rich clusters at redshifts above 1.5, which would be qualitatively consistent with certain findings in group-level overdensities at lower redshifts \citep{sob11}.  This is most likely explained by a short time spent in the clusters, less than the delay time for quenching to show its effects on galaxy colors and emission lines, and/or dynamical immaturity of the clusters.  More detailed research of galaxy properties in clusters at these redshifts, such as indicators of morphological changes that may precede quenching, or spectroscopic confirmation of passive galaxies with sensitive near-IR spectroscopy on the next generation of telescopes and instruments to allow for dynamical analysis of the clusters, may help clarify the situation.

\section*{Acknowledgements}

This research uses observations obtained with MegaPrime/MegaCam, a joint project of CFHT and CEA/DAPNIA, at the Canada-France-Hawaii Telescope (CFHT) which is operated by the National Research Council (NRC) of Canada, the Institut National des Sciences de l'Univers of the Centre National de la Recherche Scientifique (CNRS) of France, and the University of Hawaii. This work is based in part on data products produced at TERAPIX and the Canadian Astronomy Data Centre as part of the Canada-France-Hawaii Telescope Legacy Survey, a collaborative project of NRC and CNRS. This project also uses observations taken at the ESO Paranal Observatory (ESO programs 085.A-0166, 085.A-0613, 086.A-0398, 087.A-0145, 087.A-0483, 088.A-0639, 089.A-0125, and 091.A-0478), Las Campanas Observatory in Chile, the Keck Telescopes in Hawaii, and the Spitzer Space Telescope, which is operated by the Jet Propulsion Laboratory, California Institute of Technology under a contract with NASA.

Our research team recognizes support from the following grants: Universidad Andres Bello internal research project DI-12-19/R (JN), BASAL Center for Astrophysics and Associated Technologies (CATA) grant AFB-170002 (RD), and ALMA-CONICYT grant no 31180051 (PC).  
This work was supported in part by NSF grants AST-1517863, AST-1518257, and AST-1815475.
Additional support was provided by NASA through grants GO-15294 and AR-14289 from the Space Telescope Science Institute, which is operated by the Association of Universities for Research in Astronomy, Inc., under NASA contract NAS 5-26555. 
Support is acknowledged from a European Space Agency (ESA) research fellowship (LJO) and by grant number 80NSSC17K0019 issued through the NASA Astrophysics Data Analysis Program (ADAP).

\section*{Data Availability}

The near-IR spectroscopy used in this paper is available to the public on the Keck Observatory Archive (KOA) Data Access Service: https://koa.ipac.caltech.edu/cgi-bin/KOA/nph-KOAlogin.  Information about the nights of observation is available in the Appendix. To request access to other data, please ask the authors.



\bibliographystyle{mnras}
\bibliography{sfms} 




\appendix

\section{Summary of Observation}
\begin{table*}
	\caption{Spectroscopic Data Summary}
	\label{tab:obs}
	\centering
	\begin{tabular}{lcccccc}
		\hline\hline
		Cluster & Mask & Date & $N_{obj}$ & $N_{1D}$ & $T_{exp,tot}$ & $N_{exp}$   \\ 
		& & & & & min. &  \\ 
		\hline
		SpARCS0330 & 1 & 2012 Nov 25 & 40 & 40 & 37.8 & 19 \\
		SpARCS0330 & 2 & 2012 Nov 25 & 33 & 33 & 39.8 & 20 \\
		SpARCS0330 & 3 & 2012 Nov 25 & 35 & 35 & 39.8 & 20 \\
		SpARCS0330 & 4 & 2012 Nov 25 & 41 & 41 & 23.9 & 12 \\
		SpARCS0330 & 5 & 2014 Nov 16 & 40 & 40 & 95.4 & 48 \\
		SpARCS0330 & 6 & 2014 Nov 16 & 36 & 36 & 111.3 & 56 \\
		SpARCS0330 & 7 & 2014 Nov 16 & 40 & 40 & 71.6 & 36 \\
		SpARCS0330 & 8 & 2014 Nov 17 & 40 & 40 & 95.4 & 48 \\
		SpARCS0224 & 1 & 2012 Nov 25 & 38 & 38 & 35.8 & 18 \\
		SpARCS0224 & 2 & 2012 Nov 25 & 38 & 37 & 39.8 & 20 \\
		SpARCS0224 & 3 & 2012 Nov 25 & 38 & 35 & 19.9 & 10 \\
		SpARCS0224 & 6 & 2014 Aug 16 & 28 & 28 & 79.5 & 40 \\
		SpARCS0224 & 7 & 2014 Aug 16 & 29 & 29 & 77.5 & 39 \\
		SpARCS0224 & 8 & 2014 Nov 16 & 36 & 36 & 87.5 & 44 \\
		SpARCS0225 & 1 & 2014 Nov 17 & 35 & 35 & 61.6 & 31 \\
		SpARCS0225 & 2 & 2014 Nov 17 & 39 & 39 & 63.6 & 32 \\
		\hline  
	\end{tabular}
\end{table*}
Table A1 summarizes the exposures, providing the dates (column 3), numbers of objects observed (column 4), numbers of objects with successful 1-D extractions (column 5), total exposure times in minutes (column 6), exposure time for an individual observation (column 7), and total number of exposures of the mask for all of our successful observations (column 8).  Total exposure times range from 20 minutes to 111 minutes, but most were between 35 and 90 minutes.


\bsp	
\label{lastpage}
\end{document}